\documentclass[iop,twocolumn]{emulateapj}
\slugcomment{Draft, \today}

\usepackage{savesym}
\usepackage{natbib}

\usepackage{amsmath,amssymb,bm,stmaryrd,amsthm,mathrsfs,dsfont}
\usepackage{multirow}
\usepackage{graphicx}
\usepackage{subfigure}
\usepackage{captcont}
\usepackage{float}
\usepackage{booktabs}
\usepackage{epstopdf}
\usepackage{color}

\graphicspath{{figures/}}

\def\be{\begin{equation}}
\def\ee{\end{equation}}
\def\kms{{\rm \,km\,s^{-1}}}
\def\Gyr{{\rm \,Gyr}}

\def\kpc{{\rm \,kpc}}

\def\eV{{\rm \,eV}}
\def\keV{{\rm \,keV}}
\def\K{{\rm \,K}}

\def\msun{{\,M_\odot}}
\def\dh70{\,h_{70}^{-1}}
\def\ghz{\rm \,GHz}

\newcommand\na{\rmfamily{New A}}

\begin{document}

\title{Simulating the galaxy cluster ``El Gordo'': gas motion, kinetic Sunyaev-Zel'dovich signal, and X-ray line features}
\shorttitle{Expected observational signatures of gas motion in ACT-CL J0102--4915}
\shortauthors{Zhang, Yu, \& Lu}
\author{Congyao Zhang$^{1,2}$, Qingjuan Yu$^{2,\dagger}$, and Youjun Lu$^{3,4}$}
\affil{
$^1$~Max Planck Institute for Astrophysics, Karl-Schwarzschild-Str. 1, D-85741 Garching, Germany \\
$^2$~Kavli Institute for Astronomy and Astrophysics, Peking University, Beijing, 100871, China; $^\dagger$yuqj@pku.edu.cn \\
$^3$~National Astronomical Observatories, Chinese Academy of Sciences, Beijing, 100012, China\\
$^4$~School of Astronomy and Space Sciences, University of
Chinese Academy of Sciences, No. 19A Yuquan Road, Beijing 100049, China
}

\begin{abstract}
The massive galaxy cluster ``El Gordo'' (ACT-CL J0102--4915) is a rare merging
system with a high collision speed suggested by multi-wavelength observations
and the theoretical modeling. \citet{Zhang2015} propose two types of mergers,
a nearly head-on merger and an off-axis merger with a large impact parameter,
to reproduce most of the observational features of the cluster, by using
numerical simulations. The different merger configurations of the
two models result in different gas motion in the simulated clusters. In this
paper, we predict the kinetic Sunyaev-Zel'dovich (kSZ) effect, the relativistic
correction of the thermal Sunyaev-Zel'dovich (tSZ) effect, and the X-ray
spectrum of this cluster, based on the two proposed models. We find that (1)
the amplitudes of the kSZ effect resulting from the two models are both on
the order of $\Delta T/T\sim10^{-5}$; but their morphologies are different,
which trace the different line-of-sight velocity distributions of the systems; (2) the
relativistic correction of the tSZ effect around $240\ghz$ can be possibly used
to constrain the temperature of the hot electrons heated by the shocks; and
(3) the shift between the X-ray spectral lines emitted from different regions
of the cluster can be significantly different in the two models. The shift and
the line broadening can be up to $\sim 25{\rm\,eV}$ and $50{\rm\,eV}$, respectively.
We expect that future observations of the kSZ effect and the X-ray spectral
lines (e.g., by ALMA, XARM) will provide a strong constraint on the gas
motion and the merger configuration of ACT-CL J0102--4915.

\end{abstract}


\section{Introduction} \label{sec:introduction}

Major mergers of galaxy clusters with relative velocity up to thousands of
$\kms$ can lead to violent disturbances in the density and the velocity fields
of the intracluster medium (ICM). The line-of-sight motion of the gas in the
ICM could be measured by the Sunyaev-Zel'dovich (SZ) signal via the kinetic SZ
(kSZ) effect (see a review in \citealt{Birkinshaw1999}), and by the X-ray
observations via the Doppler shifting and broadening of the spectral lines
\citep{Sunyaev2003,Biffi2013}. These measurements are expected to
provide insights into understanding the cluster merger configurations, the
cosmic velocity field, and the gastrophysics of the ICM, though being a great
challenge for observational techniques.

Over the past years, several efforts have been made to detect the kSZ signals
via the statistical stacking method
\citep{Hand2012,Schaan2015,Planck2016_I37,Soergel2016} or observing the kSZ
effect directly towards individual galaxy clusters
\citep{Mroczkowski2012,Sayers2013,Sayers2016}. One of the scientific goals for
the present and the future cosmic microwave background (CMB) experiments is to
map the kSZ effect resulting from large-scale structures
\citep[e.g.,][]{Benson2014}. On the other hand, the new generation of X-ray
instruments with unprecedented high spectral resolution is being developed to
reveal the nature of gas motions in galaxy clusters,
e.g., \emph{XARM} (also known as \emph{ASTRO-H2}), \emph{ATHENA}.

In this work, we aim to predicting the observational features of the kSZ effect
and the X-ray spectral lines of massive galaxy cluster ACT-CL J0102--4915 (``El
Gordo'') by using simulations. ACT-CL J0102--4915, first discovered by the
Atacama Cosmology Telescope (ACT, see \citealt{Marriage2011}), is a rare
merging system with a high collision speed \citep{Menanteau2012,Jee2014}. Its
unique observational characteristics and inferred merging configurations make
it an ideal target to investigate the motions of ICM during the cluster merging
process. \citet[][hereafter ZYL15]{Zhang2015} have numerically explored
the merger scenario of ACT-CL J0102--4915, and proposed two models that can
reproduce most of the features revealed by observations. In this paper, we present the
simulated kSZ effect and the X-ray spectra for the two fiducial models in
ZYL15, and show that future observations on these two aspects may help to
constrain the merger configuration of ACT-CL J0102--4915.

This paper is organized as follows. In Section~\ref{sec:method}, we describe
the method of simulating ACT-CL J0102--4915 and modeling its secondary CMB
anisotropies and X-ray spectra. Our main results, including the kSZ spatial
distributions, the SZ spectra, relativistic correction of the tSZ effect,
and the X-ray spectral lines for both of the fiducial models, are presented
in Section~\ref{sec:results}. In Section~\ref{sec:conclusions}, we summarize
our conclusions.

\section{Method} \label{sec:method}

\subsection{The simulations} \label{sec:method:simulations}

ZYL15 have carried out a detailed study on identifying the merger
configuration of ACT-CL J0102--4915 by using extensive numerical simulations, and found
two fiducial models (A and B) that can match most of the observations of ACT-CL
J0102--4915. Before describing the methodology used in this paper, we first
summarize the simulations that have been performed in ZYL15 in this section.

ZYL15 perform simulations of mergers between two ideal galaxy clusters
by adopting both the GADGET-2 code \citep{Springel2001} and the FLASH code
\citep{Fryxell2000, Ricker2008}. Each cluster is simplified as a spherical halo
consisting of collisionless dark matter and collisional gas, and
the gas is assumed to be adiabatic. The merger parameters of the
initial conditions include the mass of the primary (secondary) cluster
$M_1\,(M_2)$, the mass ratio $\xi\,(\equiv M_1/M_2)$, the gas fraction of the
primary (secondary) cluster $f_{\rm b1}\,(f_{\rm b2})$, the initial relative
velocity $V$, and the impact parameter $P$. The parameter settings of fiducial
models A and B are summarized in Table~\ref{tab:ic_para}. These two models
have different merger configurations: one is a nearly head-on merger with
$P=300\dh70\kpc$ (model A); the other is a highly off-axis merger with
$P=800\dh70\kpc$ (model B).  These two models match the
observations best (including the X-ray luminosity and the temperature
distributions, the SZ temperature decrement, the Mach numbers, etc.; see Table~2
in ZYL15) at $t=0.13$ and $0.14\Gyr$ after the first pericentric passage,
respectively. Model B can generate a remarkable wake-like substructure in
the X-ray surface brightness map, like the one seen in the \textit{Chandra}
observation \citep{Menanteau2012}. Moreover, model B presents a larger
radial relative velocity ($V_{\rm r}=1820\kms$) than model A
($V_{\rm r}=960\kms$), which, together with different impact parameters,
is expected to be the main reason why these two models show different
kSZ effects (see details in Section~\ref{sec:results:secondary_cmb}).

An extended model B, whose initial merger conditions are mostly the same as
those in model B, is further proposed in ZYL15 to explore the effects of the
gas density profile on the observed X-ray emission at the cluster outskirts.
For simplicity, we focus on describing the
effects of the above two fiducial models in this work, which is sufficient to
illustrate the difference caused by their different impact parameters and
radial velocities as to be seen below.  Together with the results of the
fiducial models, the results of extended model B will also be shown in the
figures of this paper, which do not affect the main conclusion drawn from
the comparison of the fiducial models.

In Section~\ref{sec:method:mock_obs}, we describe the method to model the
secondary CMB anisotropies and the X-ray spectra for models A and B.
The GADGET-2 and the FLASH simulations show little difference in the results
for the purpose of this work.  Unless specifically stated, all of the results
below are based on the FLASH simulations presented in ZYL15.

\begin{table*}
\begin{center}
  \caption{Initial merger parameters}
  \label{tab:ic_para}
\begin{tabular}{cccccc}
  \hline\hline
  & $M_1\ (\msun)$ & $\xi$ & $V\ (\kms)$ & $P\ (\dh70\kpc)$ & $(f_{\rm b1},\,f_{\rm b2})$ \\
  \hline
  Model A & $1.3\times10^{15}$ & $2.0$ & $3000$ & $300$ & (0.10,\,0.10) \\
  Model B & $2.5\times10^{15}$ & $3.6$ & $2500$ & $800$ & (0.05,\,0.10) \\
  Extended Model B & $2.5\times10^{15}$ & $3.6$ & $2500$ & $800$ & (0.11,\,0.12) \\
  \hline\hline
\end{tabular}
\end{center}
\end{table*}

\subsection{Mocking the secondary CMB anisotropies and the X-ray spectra} \label{sec:method:mock_obs}

For any given merger snapshots, we can generate the maps of the secondary CMB
anisotropies resulting from the cluster in the observer's sky plane,
including the tSZ effect, the kSZ effect, and the Rees-Sciama (RS)
effect by the following equations. In these equations, the subscripts ``$||$''
and ``$\perp$'' of a (position or velocity) vector are used to represent its
components parallel and perpendicular to the line of sight, respectively.
\begin{itemize}
\item The changes in the CMB temperature at frequency $\nu$ by the tSZ effect
can be obtained from the gas mass density $\rho_{\rm gas}(\textbf{r})$ and the
temperature $T_{\rm gas}(\textbf{r})$ distributions by
  \be
  \frac{\Delta T}{T_{\rm CMB}}\Big|^\nu_{\rm tSZ}  =  \frac{\sigma_{\rm
T}k_{\rm B}}{m_{\rm e}c^{2}} \int n_{\rm e}(\textbf{r}) T_{\rm gas}(\textbf{r})
\mathcal{F}(x_\nu,\Theta){\rm d}\textbf{r}_{\parallel},
  \label{eq:tsz}
  \ee
  where $n_{\rm e}(\textbf{r})\equiv 0.9\rho_{\rm gas}(\textbf{r})/m_{\rm H}$
is the electron number density, and $T_{\rm CMB},\ \sigma_{\rm T},\ k_{\rm B},\
m_{\rm e},\ m_{\rm H}$, and $c$ represent the CMB temperature, the Thomson
cross section, the Boltzmann constant, the electron mass, the hydrogen atom
mass, and the speed of light, respectively. The tSZ distortion $\mathcal{F}(x_\nu,\Theta)$
is a function of $x_\nu\,(\equiv h\nu/k_{\rm B}T_{\rm CMB})$ and $\Theta\,
(\equiv k_{\rm B}T_{\rm gas}/m_{\rm e}c^{2})$. For the polynomial approximation,
$\mathcal{F}(x_\nu,\Theta)=\sum^{i=4}_{i=0}{Y_{i}(x_\nu)\Theta^{i}}$ includes
the relativistic correction (see eqs.~2.25-2.30 in \citealt{Itoh1998}), which,
however, becomes less convergent when the gas temperature is higher
($\geq30\keV$, see more discussions in \citealt{Itoh1998} and \citealt{Nozawa2000}).
The shock-heated gas in the simulations can reach a temperature higher than $50\keV$
(see top panels in Fig.~\ref{fig:xray_spec}). To precisely determine the tSZ
effect (particularly around the crossover frequency, $\nu_0$, defined by the
frequency at which the tSZ effect vanishes), we adopt the numerical
solution of the Boltzmann equation (see eqs.~2.1-2.5 in \citealt{Itoh1998})
as $\mathcal{F}(x_\nu,\Theta)$ in Equation~(\ref{eq:tsz}). It is worth noting
that our simulations model the ICM by using a single gas component, which
implicitly assumes the electrons and ions reach the thermal equilibrium
instantly. However, in reality this might not be the case. The electron--ion
thermal equilibrium time--scale in the ICM is still poorly constrained
(see \citealt{Russell2012} and references therein). The tSZ effect at
the shock region obtained from our simulations would be overestimated if the
transport process between the electrons and ions is not sufficiently efficient.
On the other hand, we find that the relativistic correction
of the tSZ effect may be useful to give a constraint on the temperature of the
shock-heated electrons, as to be demonstrated in Section~\ref{sec:results:secondary_cmb}.
\item The changes in the CMB temperature at frequency $\nu$ by the kSZ effect
is determined by
  \begin{align}
  \frac{\Delta T}{T_{\rm CMB}}\Big|^\nu_{\rm kSZ} & = \frac{\sigma_{\rm
T}k_{\rm B}}{m_{\rm e}c^{2}} \int n_{\rm e}(\textbf{r}) \times  \nonumber \\ &
\Big\{-{\bm\beta}_{||}\cdot({\textbf{r}}_{\parallel}/|{\textbf{r}}_{\parallel}|)\Big[1+C_{1}(x_\nu)\Theta+C_{2}(x_\nu)\Theta^{2}\Big]+ \nonumber \\ &
{|\bm\beta|}^2\Big[\frac{1}{3}Y_{0}(x_\nu)+\Big(\frac{5}{6}Y_{0}(x_\nu)+\frac{2}{3}Y_{1}(x_\nu)\Big)\Theta\Big]+
\nonumber \\ &
\frac{1}{2}(3|{\bm\beta}_{||}|^2-|{\bm\beta}|^2)\Big[D_{0}(x_\nu)+D_{1}(x_\nu)\Theta\Big]\Big\}{\rm
d}\textbf{r}_{\parallel},
  \label{eq:ksz} \end{align}
  where ${\bm\beta}\ (\equiv \textbf{v}/c)$ and ${\bm\beta}_{||}\ (\equiv
\textbf{v}_{||}/c)$ are the parameterized gas velocities relative to the
peculiar velocity of the center of mass of the cluster,
${\textbf{r}}_{\parallel}/|{\textbf{r}}_{\parallel}|$ is the unit vector of the
line of sight directing away from us; $Y_{0}(x_\nu)$,
$Y_{1}(x_\nu)$, $C_{1}(x_\nu)$, $C_{2}(x_\nu)$, $D_{0}(x_\nu)$, and
$D_{1}(x_\nu)$ are the coefficients for the polynomial approximation of the
relativistic correction (see eqs.~24--32 in \citealt{Nozawa1998}). Since the
analytic expression of the kSZ effect (i.e., Eq.~\ref{eq:ksz}) can provide a
convergent result in our study, we simply adopt this form for the calculation.
Note here that Equation~(\ref{eq:ksz}) does not involve the kSZ effect induced
by the peculiar motion of the center of mass of the cluster.
\citet{Lindner2015} presents a best-fit peculiar velocity of ACT-CL J0102--4915
as $v_p=-1100^{+1300}_{-2200}$ (or $-2800^{+1700}_{-3100}$)$\kms$ by using three
(or five) bands observations of the SZ spectra. However, in this work we still
assume that the center of mass of the entire merging system is at rest in the CMB
rest frame (i.e., $v_p=0$), for the following reasons. (1) The main scientific goal
of this paper is to investigate and predict the observational features of the gas motion
caused by the merging process. (2) The observational constraints on $v_p$ still
have large uncertainties (see \citealt{Lindner2015}). (3) We find that the kSZ
effect $\mathcal{R}(v_p,\,x_\nu,\Theta)$ induced by the peculiar velocity of
the cluster could be well modeled by
  \be
  \mathcal{R}(v_p,\,x_\nu,\Theta) \simeq
\kappa\cdot\frac{v_p}{1000\kms}\cdot\frac{\Delta T}{T_{\rm
CMB}}\Big|^{\nu=148\ghz}_{\rm tSZ},
  \ee where $\kappa=0.085$ for both of models A and B, which does not significantly
   depend on $\nu$ within the frequency range we are interested in. Thus, it could be
   easily separated from the other SZ components.
\item The CMB temperature changes caused by the RS effect could be approximated as
  \be
  \frac{\Delta T}{T_{\rm CMB}}\Big|_{\rm RS} = \frac{4G}{c^3}\int{
  {\rm d}^3\textbf{r}'\rho_{\rm tot}(\textbf{r}')\textbf{v}(\textbf{r}')
  \cdot\frac{\textbf{r}_{\perp}-\textbf{r}_{\perp}'}{|\textbf{r}_{\perp}-\textbf{r}_{\perp}'|^2}},
  \ee
  where $G$ represents the gravitational constant; $\rho_{\rm tot}(\textbf{r})$
is the cluster total mass density (see eq.~4 in \citealt{Martin2004}).
\end{itemize}
We present the results of the secondary CMB anisotropies at $\nu=148$ and
$218\ghz$ in Section~\ref{sec:results:secondary_cmb} to match the frequencies
covered by the current high resolution SZ measuring instruments, e.g., ACT,
South Pole Telescope (SPT).

We model the X-ray spectra of ACT-CL J0102--4915 by
\be
S_{\rm X}=\frac{1}{4\pi (1+z)^3\Delta\nu_0}\int n_{\rm e}(\textbf{r})n_{\rm
H}(\textbf{r})\Lambda_{\nu}(T_{\rm gas},\, Z) {\rm d}\nu_0{\rm
d}\textbf{r}_{||},
\label{eq:xray}
\ee where $\Lambda_{\nu}(T_{\rm gas},\, Z)$ is the cooling function at the
frequency $\nu=\nu_0\sqrt{(1+|{\bm\beta}_{||}|)/(1-|{\bm\beta}_{||}|)}$,
obtained by adopting the MEKAL model in the XSPEC v12.8 package
\footnote{See
\href{http://heasarc.nasa.gov/xanadu/xspec/}{http://heasarc.nasa.gov/xanadu/xspec/}}.
Here $n_{\rm H}(\textbf{r})\equiv 0.83n_{\rm e}(\textbf{r})$ is the number
density of the hydrogen; the spectral resolution $\Delta\nu_0$ is set to $5
{\rm\,eV}$.  We further assume a uniform metallicity distribution
($Z=0.3Z_\odot$) of the cluster in the calculation, where the solar metal
abundance $Z_\odot$ is adopted from \citet{Anders1989}. In this study, we are
particularly interested in the H-like ($\sim7.0\keV$) and He-like
($\sim6.7\keV$) Fe-K lines emitted from the hot ICM. To model the thermal
broadening of the spectral lines due to the thermal motion of the ions, we
simply smooth the cooling function by a Gaussian kernel with the full width at
half maximum (FWHM) of $4.9{\rm \,eV}(k_{\rm B}T_{\rm gas}/5\keV)^{1/2}$
(see eq.~11 in \citealt{Kitayama2014}). In ACT-CL J0102--4915, the high
line-of-sight bulk velocities of the gas play the dominant role in the
broadening of the X-ray spectral lines (see
Section~\ref{sec:results:xray_spectra}).

\section{Modeling Results} \label{sec:results}

\subsection{The secondary CMB anisotropies} \label{sec:results:secondary_cmb}

\begin{figure*} 
\centering
\includegraphics[width=0.9\textwidth]{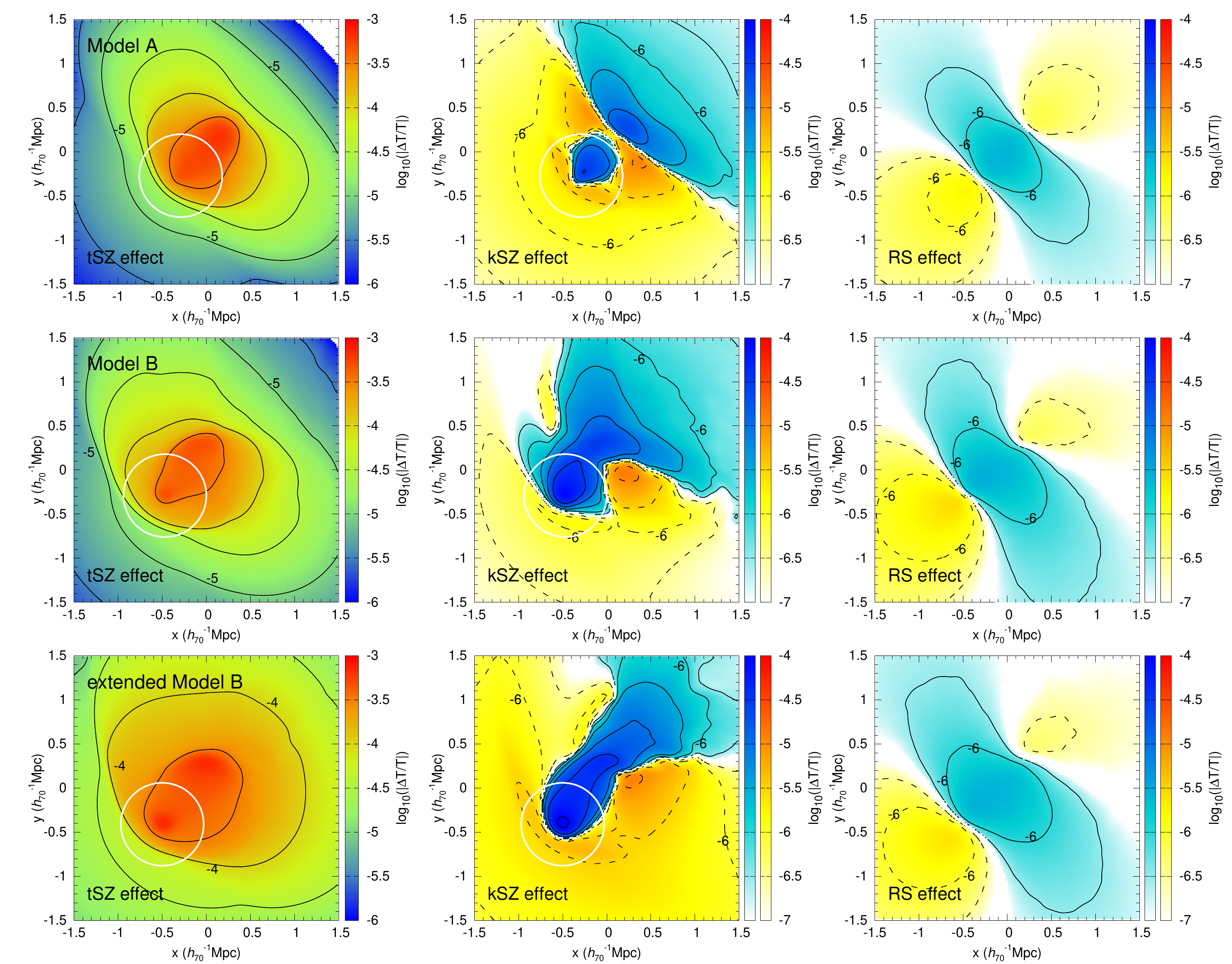}
\caption{The distributions of the tSZ effect, the kSZ effect, and the RS
effect resulting from a merging galaxy cluster with the merging
configurations in fiducial model A (top panels), fiducial model B
(middle panels), and extended model B (bottom panels).
All the maps are estimated in the $148\ghz$ band, and smoothed by a
Gaussian kernel with an FWHM of $10''$. In the second and the third
columns, the red-yellow colorbar (dashed black contour lines) and the
blue-cyan colorbar (solid black contour lines) represent the positive
and the negative CMB temperature variations, respectively.
In all the panels, the interval between two successive contour levels
with the same line types is $0.5$ in $\log_{10}(|\Delta T/T|)$.
The SZ surface brightness averaged within the white circles (with
the center being at the position of the maximum of the X-ray emission and the
radius being $1'$) are shown in Figure~\ref{fig:isz}. This figure shows a
significant difference between models A and B in the morphology of the kSZ
effect distributions (see Section~\ref{sec:results:secondary_cmb}). }
\label{fig:sz_sim}
\end{figure*}

Figure~\ref{fig:sz_sim} shows the distributions of the tSZ effect, the kSZ
effect, and the RS effect for fiducial models A (top panels) and B (middle
panels), and extended model B (bottom panels) at $148\ghz$. The images have
been smoothed by a Gaussian kernel with an FWHM of $10''$, which corresponds
to the measurement capability of the Atacama Large Millimeter/Submillimeter
Array (ALMA). Fiducial model A shows an almost axisymmetric morphology of its
secondary CMB anisotropies with respect to the merger direction, because it is
a nearly head-on merger (see Table~\ref{tab:ic_para}); while the distributions
resulting from model B look asymmetric. By comparing the panels in
Figure~\ref{fig:sz_sim}, we come to the following conclusions.
\begin{itemize} \item At $148\ghz$, the tSZ effect dominates the secondary CMB
anisotropies resulting from ACT-CL J0102--4915, which reveals the hot electron
density distributions in the merger system. Fiducial models A and B show a
generally similar morphology of the tSZ effect distributions. In model B, the
significant tSZ signals extend to a larger radius, because model B has a higher
cluster mass (i.e., $3.2\times10^{15}\msun$).
  \item The kSZ effect traces the motions of the gas component in galaxy
clusters. In the middle column of Figure~\ref{fig:sz_sim}, the red-yellow and
the blue-cyan colorbars represent the positive (i.e., most of the gas moving
towards the observer along the line-of-sight) and the negative (i.e., that
moving away from the observer) kSZ effect signals, respectively. The interfaces
between the positive and negative signals mark the positions of the shocks or
the cold fronts formed during the merger process (see also the temperature
distributions in Fig.~\ref{fig:xray_spec}). At $148\ghz$, the strength of the
kSZ effect is around one order of magnitude weaker than that of the tSZ effect in both of the models.
Fiducial model B reveals a twisted morphology of the kSZ effect, because of its
off-axis merger configuration. The maximal strength of the kSZ effect of model
B is approximately three times higher than that of model A. This is mostly due
to the higher relative radial velocity between the two merging subclusters in
model B (see Fig.~\ref{fig:xray_spec}b, and also section~3.3.5 in ZYL15 for
more discussions on the relative radial velocity of ACT-CL J0102--4915). The
differences in the morphology of the kSZ effect between models A and B may
provide a way to distinguish these two fiducial models in the future SZ
observations of ACT-CL J0102--4915.
  \item The strength of the RS effect is more than 10 times weaker than that of
the kSZ effect for both of models A and B. For the capability of the current SZ
measuring instruments (e.g., ACT or SPT), contributions of the RS effect to the
secondary CMB temperature anisotropies resulting from ACT-CL J0102--4915 are
negligible.  \end{itemize}

\begin{figure*} 
\centering
\includegraphics[width=0.7\textwidth]{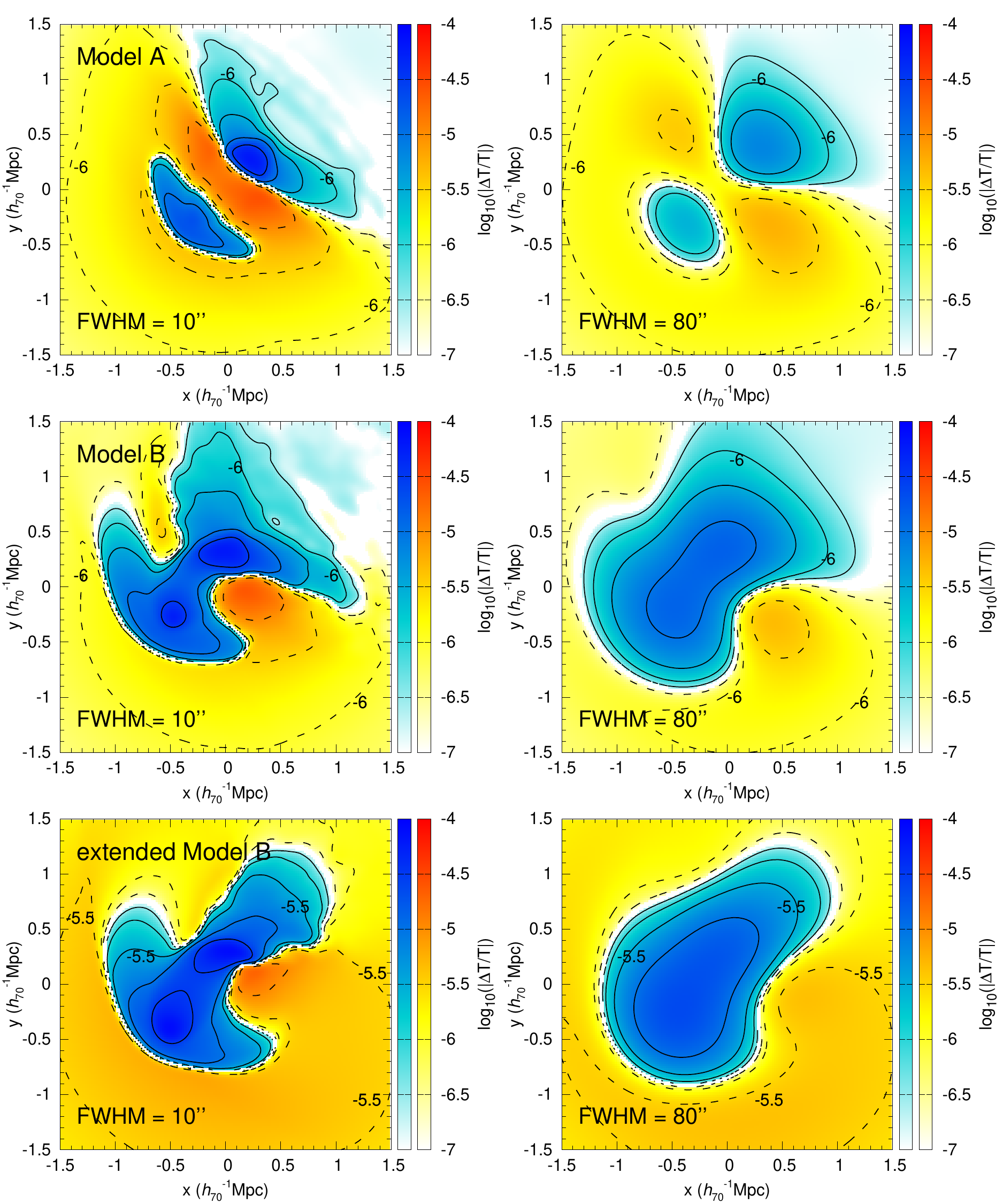}
\caption{Legends similar to those in Figure~\ref{fig:sz_sim}, but for the
extracted kSZ effect distributions by combining the SZ maps at $148$ and
$218\ghz$ (see Eq.~\ref{eq:extracted_ksz}). The results are smoothed by a
Gaussian kernel with $\rm FWHM=10''$ (left panel) and FWHM$=80''$
(right panel), respectively. This figure indicates that the measuring
sensitivity to detect the kSZ signals of ACT-CL J0102--4915 and distinguish
between the proposed merging configurations needs to be better
than $10^{-5}\K$ level (see Section~\ref{sec:results:secondary_cmb}).
}
\label{fig:sz_obs}
\end{figure*}

In the Kompaneets approximation, the crossover frequency of the tSZ effect is
$\simeq218\ghz$, which provides a unique window for detecting the kSZ signals.
However, the relativistic correction for the gas with high temperature
contributes non-negligible tSZ signals to the CMB map at $218\ghz$. The
strengths of the thermal and the kinetic SZ effects of ACT-CL J0102--4915 are
actually comparable at $218\ghz$. To extract the kSZ signals, we combine the
distributions of the SZ effect at $148$ and $218\ghz$ as follows,
\be \frac{\Delta T}{T_{\rm CMB}}\Big|_{\rm res}=\frac{1}{1-\alpha}\Big(\frac{\Delta T}{T_{\rm
CMB}}\Big|^{218\ghz}_{\rm tSZ+kSZ}-\alpha\cdot \frac{\Delta T}{T_{\rm
CMB}}\Big|^{148\ghz}_{\rm tSZ+kSZ}\Big),
\label{eq:extracted_ksz} \ee
where $\alpha$ is a parameter selected to maximally remove the contamination of
the tSZ effect in the residual image\footnote{We obtain the best-fit parameter
$\alpha$ by minimizing the summation of
${\Big(\frac{\Delta T}{T_{\rm CMB}}\Big|^{218\ghz}_{\rm tSZ}-\alpha\cdot
\frac{\Delta T}{T_{\rm CMB}}\Big|^{148\ghz}_{\rm tSZ}\Big)^2}$ over all the
pixels of the thermal SZ images.}.
The best-fit parameter $\alpha$ is $0.143$ and $0.154$ for fiducial models A
and B, respectively. As a reference, the ratio of the tSZ distortions
at $218\ghz$ and $148\ghz$ is $\mathcal{F}(x_{\nu=218\ghz},\Theta)/
\mathcal{F}(x_{\nu=148\ghz},\Theta)=0.095$ (see Eq.~\ref{eq:tsz}), when fixing
the gas temperature to the best-fit cluster X-ray temperature ($\sim15\keV$).
The best-fit parameter $\alpha$ is larger than this value so that
some tSZ signal produced by the high-temperature gas can be removed.
It is worthy of noting that, in reality, the additional contaminations in the
cm-wavelength band (e.g., dusty galaxies, radio point sources) may make the
situation more complicated.

The recovered kSZ effect residual maps are presented in
Figure~\ref{fig:sz_obs}. The distributions (left panels) are generally
in agreement with the models shown in the middle column of
Figure~\ref{fig:sz_sim}, which illustrate that most of the tSZ emissions
have been effectively removed after the extracting process.
However, the tSZ contamination contributed by the high-temperature gas can
still be seen around the cluster shock regions (see also left panels in
Fig.~\ref{fig:sz_rc}). By comparing the residual images at different
angular resolutions in Figure~\ref{fig:sz_obs}, we find that even though
the low-resolution ones (right panels with FWHM $\sim1.3'$)
lack substructures and high signal contrasts, the differences between
fiducial models A and B are still distinguishable in these maps.
Our results suggest that, to detect the kSZ signal of ACT-CL J0102--4915
in the future, the CMB measurement sensitivity better than $10^{-5}\K$ is
needed. ALMA should have the capability to reach such a goal
(see \citealt{Basu2016,Yamada2012}).
It may be harder to extract the kSZ effect of ACT-CL J0102--4915
in reality than in the illustration here, because no further observational
contaminations and uncertainties are involved in our models.

\begin{figure} 
\centering
\includegraphics[width=0.45\textwidth]{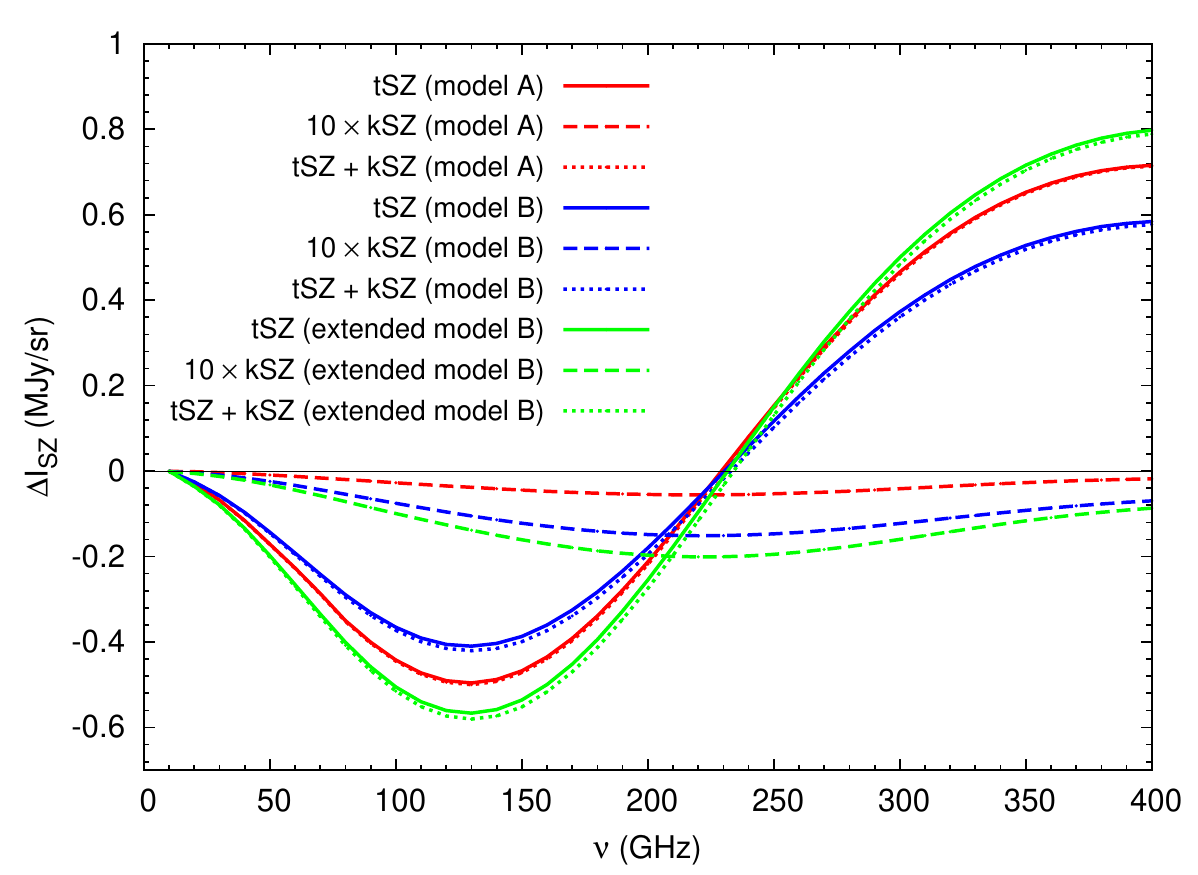}
\caption{Spectral distortion of the CMB radiation due to the tSZ effect, the
kSZ effect, and their combination. The vertical axis shows the average intensity
distortion within the white circles marked in Figure~\ref{fig:sz_sim}. To present
a clear comparison with the tSZ spectra, we show the kSZ spectra amplified by a
factor of $10$ as indicated in this figure. The strength of the kSZ spectra is
around $1/30$--$1/20$ of that of the tSZ spectra. See details in
Section~\ref{sec:results:secondary_cmb}.}
\label{fig:isz}
\end{figure}

Figure~\ref{fig:isz} shows the SZ spectra of fiducial models A and B, and extended model B averaged
within the white circles indicated in Figure~\ref{fig:sz_sim}. The amplitude of
the tSZ spectrum of model A is around $20\%$ higher than that of model B,
because the core of the secondary cluster in model A shows a higher gas
temperature (see panels (a) in Fig.~\ref{fig:xray_spec} below, and also fig.~7
in ZYL15). For both of models A and B, the integrated strength of the kSZ
spectra is much weaker ($<1/20$) than that of the tSZ spectra.
We emphasize here again that the calculation does not consider the kSZ effect
induced by the peculiar velocity of the cluster.

\begin{figure} 
\centering
\includegraphics[width=0.45\textwidth]{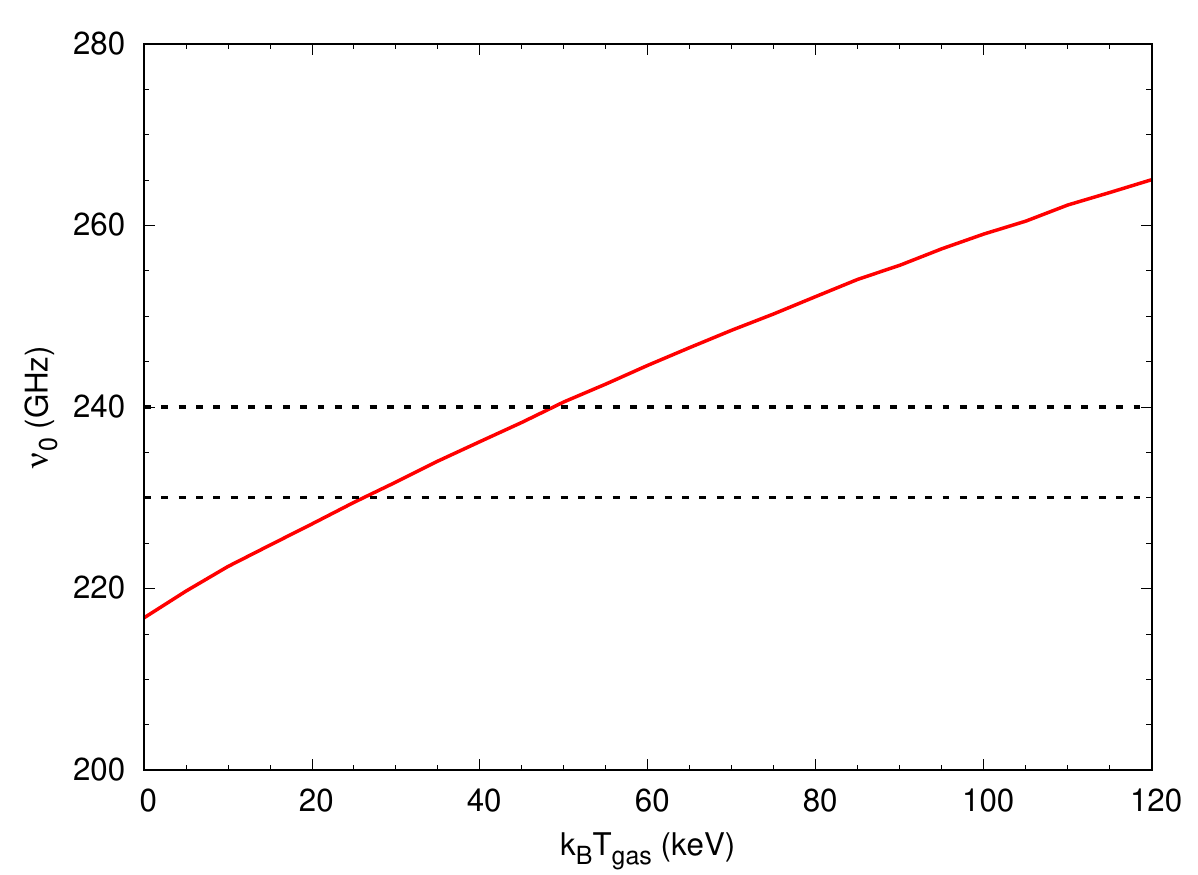}
\caption{Crossover frequency of the tSZ effect, $\nu_0$, as a function of the gas temperature.
The frequency band $\nu\in[230\ghz,\ 240\ghz]$ adopted in the right panels of
Figure~\ref{fig:sz_rc} is marked by the horizontal dashed lines. The crossover frequency
shifts to a larger value when the electron temperature gets higher. See Sections~\ref{sec:method:mock_obs} and \ref{sec:results:secondary_cmb}.}
\label{fig:zeropoint}
\end{figure}

\begin{figure*} 
\centering
\includegraphics[width=0.7\textwidth]{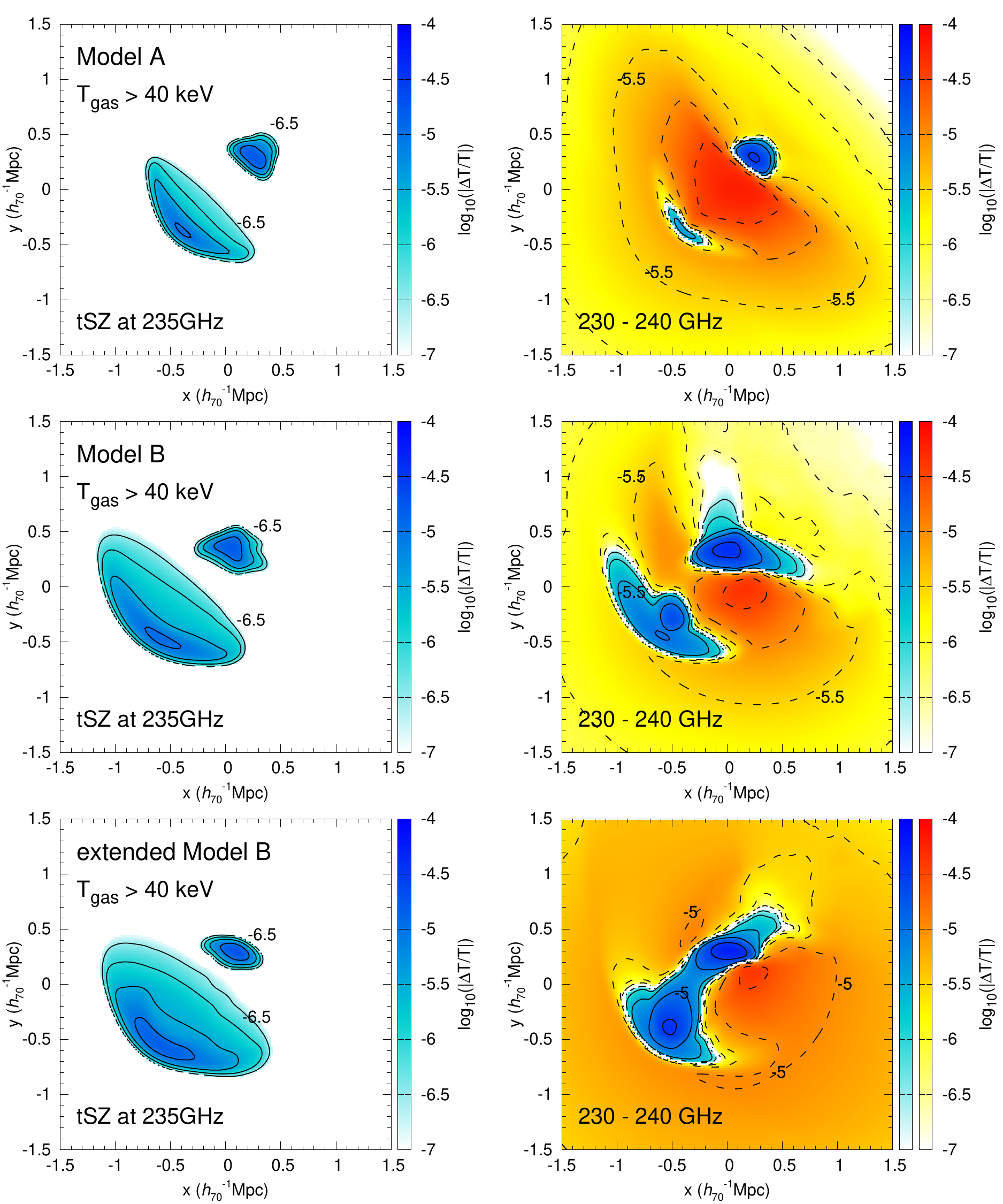}
\caption{\textit{Left panels:} tSZ effect generated by the high temperature
gas ($T_{\rm gas}>40\keV$) at $235\ghz$, which is smoothed by a Gaussian
kernel with $\rm FWHM=10''$ (the same as the right panels).
\textit{Right panels:} mock SZ observations within the frequency band
$\nu\in[230\ghz,\ 240\ghz]$ (see Eq.~\ref{eq:bandwidth}). This figure
illustrates the possibility to detect the high temperature electrons heated
by the shocks during the merger by using the relativistic correction of the tSZ effect. }
\label{fig:sz_rc}
\end{figure*}

In the last part of this subsection, we discuss the possibility
to detect the high-temperature electrons in the shock regions by using the
relativistic correction of the tSZ effect. Figure~\ref{fig:zeropoint} shows
the crossover frequency of the tSZ effect as a function of the electron
temperature. The crossover frequency increases as the electron temperature
becomes higher. At $235\ghz$, the gas hotter than $40\keV$ generates negative
tSZ signals, but the cooler gas generates the positive ones. The left panels of
Figure~\ref{fig:sz_rc} show the tSZ effect generated by the hot gas ($>40\keV$)
of models A and B at $235\ghz$. The signals coincide with the north-western (NW)
and the south-eastern (SE) shocks in positions (see top panels in
Fig.~\ref{fig:xray_spec}), and show a similar strength as the kSZ effect
(see middle column in Fig.~\ref{fig:sz_sim}). Therefore, we expect that the
tSZ effect contributed by the hot electrons in the shock regions of
ACT-CL J0102--4915 is negative if the electrons could be efficiently heated by
the ions in reality. Although being challenging, the following points should be
considered to detect this feature in future observations: (1) the electrons
are heated to high temperatures around the shock fronts, which needs high-resolution
observations ($\lesssim 10\arcsec$) to resolve; (2) the tSZ effect
contributed by the relativistic correction partly overlaps with the kSZ emission,
and they both show the negative temperature deviation and need to be isolated
properly; and (3) the strength of the relativistic correction
of the tSZ effect is sensitive to the frequency. Only the measurement
within a narrow frequency band is expected to directly see the signal.
The right panels of Figure~\ref{fig:sz_rc} show the mock SZ measurements
of models A and B within a finite band width $\nu\in[\nu_{\rm min},\ \nu_{\rm max}]$,
i.e., $\mathcal{F}$ in Equation~(\ref{eq:tsz}) is replaced by
\be
\mathcal{F}(x_{[\nu_{\rm min},\,\nu_{\rm max}]},\Theta)=\frac{\int_{\nu_{\rm min}}^{\nu_{\rm max}}\mathcal{F}(x_{\nu'},\Theta){\rm d}\nu'}{\nu_{\rm max}-\nu_{\rm min}},
\label{eq:bandwidth}
\ee
where $\nu_{\rm min}=230\ghz$ and $\nu_{\rm max}=240\ghz$ (see Eq.~\ref{eq:tsz}).
The kSZ effect is estimated at 235 GHz, since it does not significantly depend on
the frequency within this band (see Fig.~\ \ref{fig:isz}). By comparing them with
the corresponding left panels, we can see the similar pattern of the negative SZ
signal near the shock regions. The signals become insignificant if we broaden
the band width.

\subsection{The X-ray spectra} \label{sec:results:xray_spectra}

\begin{figure*} 
\centering
\includegraphics[width=0.7\textwidth]{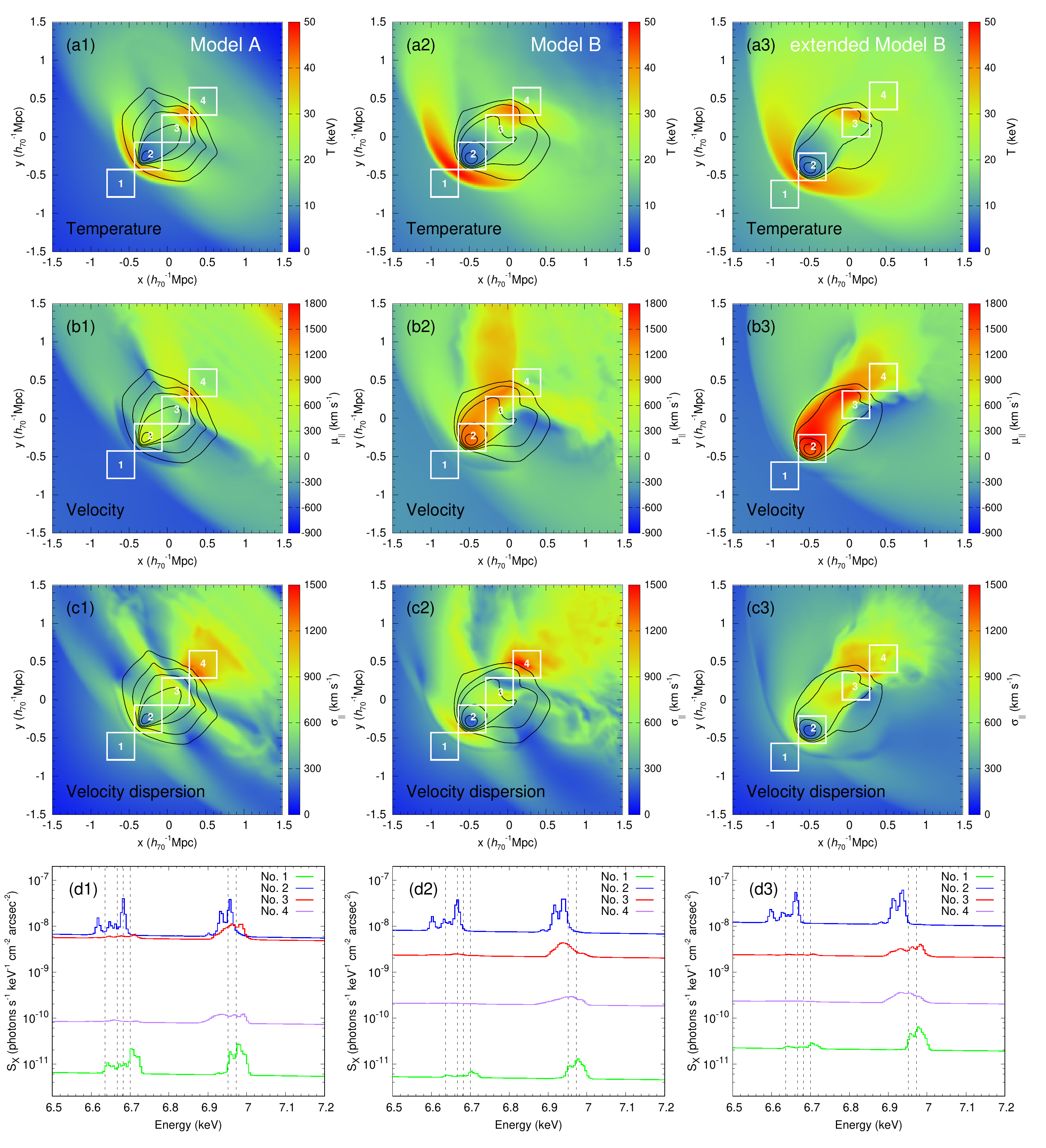}
\caption{Distributions of the spectroscopic-weighted temperature (panels a),
the line-of-sight velocity ${\mu}_{||}$ (panels b), the line-of-sight velocity
dispersion $\sigma_{||}$ (panels c), and the X-ray spectra (panels d) extracted
from the corresponding white square regions indicated in panels (a)--(c) for
fiducial model A (left column), fiducial model B (middle column), and extended
model B (right column). The spectra are presented in the cluster rest frame.
The overlaid black contours in panels (a), (b), and (c) represent the X-ray
surface brightness. The contour level starts with
$10^{-8.5}\,{\rm photons\,s^{-1}\,cm^{-2}\,arcsec^{-2}}$ and increases by a
factor of $10^{0.5}$ from outside to inside. As a reference, we mark the peak
positions of the dominant H-like and He-like Fe-K lines in the rest frame in
panels (d) by the vertical dashed lines. There is no remarkable He-like Fe-K
lines emitted from regions $3$ and $4$ in both of the models, because of the
generally high gas temperature in these areas. This figure illustrates the
remarkable shifting and the broadening of the X-ray spectral lines of the
models (see Section~\ref{sec:results:xray_spectra}).
}
\label{fig:xray_spec}
\end{figure*}

In this section, we present a detailed investigation into the broadening and
the shifting of the X-ray spectral lines of ACT-CL J0102--4915, due to the gas
motion. Before that, we examine the velocity fields of models A and B to
gain a sense of motions of the ICM caused by the cluster merging process.
Panels (a), (b), and (c) in Figure~\ref{fig:xray_spec} show the
spectroscopic-weighted temperature, line-of-sight velocity $({\mu}_{||})$, and
line-of-sight velocity dispersion $(\sigma_{||})$ distributions for fiducial
models A and B, and extended model B, respectively.
The spectroscopic-weighted temperature is defined as
\be T(\textbf{r}_{\perp}) = \frac{\int w(\textbf{r})T_{\rm gas}(\textbf{r}){\rm
d}\textbf{r}_{||}}{\int w(\textbf{r}){\rm d}\textbf{r}_{||}},
\label{eq:xray_temp} \ee
where $w(\textbf{r})=\rho_{\rm gas}^2(\textbf{r})T_{\rm
gas}^{-3/4}(\textbf{r})$ is taken as a good approximation for the projected
spectroscopic temperature of clusters obtained from the X-ray observations
\citep{Mazzotta2004}. In panels~(a), we can see the NW and SE shocks in both
of the models. The temperature of the ambient
gas has been shock-heated to around $30-50\keV$. Most of the substructures
revealed in the temperature distributions are also seen in the velocity maps
(see panels b and c), because the shocks are formed when the gas flows are
colliding with high relative velocities. Compared with the shocks in model A,
those in model B are stronger and more asymmetric, because model B has a higher
merger mass and a larger impact parameter. The shocks match the double radio
relics of ACT-CL J0102--4915 in position (see figs.~1 and 2 in
\citealt{Lindner2014}). This might be explained by that the observed radio relics is
via first-order Fermi acceleration by shocks \citep{Lindner2014}.

Panels~(b) and (c) illustrate the gas velocity distributions in these two models.
The line-of-sight velocity and velocity dispersion are defined as
\begin{align} {\mu}_{||}(\textbf{r}_{\perp}) &
=\frac{\int\epsilon(\textbf{r})\textbf{v}_{||}(\textbf{r}){\rm
d}\textbf{r}_{||}}{\int{\epsilon(\textbf{r}){\rm d}\textbf{r}_{||}}}, \\
\sigma_{||}^2(\textbf{r}_{\perp}) &
=\frac{\int\epsilon(\textbf{r})[\textbf{v}_{||}(\textbf{r})-{\mu}_{||}(\textbf{r}_{\perp})]^2{\rm
d}\textbf{r}_{||}}{\int{\epsilon(\textbf{r}){\rm d}\textbf{r}_{||}}},
\label{eq:v_moment} \end{align}
respectively, where the weight function adopts the X-ray emissivity
$\epsilon(\textbf{r})=n_{\rm e}(\textbf{r})n_{\rm
H}(\textbf{r})\int{\Lambda_{\nu}(T_{\rm gas},\,Z)}{\rm d}\nu$ covering
$0.5-7.5\keV$. As expected, the regions of the secondary gas cores show high
positive line-of-sight velocities (i.e., bulk velocities), which spatially
coincide with the cold fronts shown in the temperature distributions (i.e.,
region~2 marked in the panels). The features shown in panels (b) are similar
with those viewed in the kSZ effect distributions (see the middle column in
Fig.~\ref{fig:sz_sim}), because the amplitude of the kSZ effect is proportional
to ${\mu}_{||}$ in the first-order approximation. In panels~(c), the high
line-of-sight velocity dispersions ($500-1500\kms$) arise from (1) the
turbulence in the ICM generated during the merger, or (2) the projection of the
gas structures with different velocities along the line-of-sight. In
particular, the gas trailing after the secondary clusters in both of models A
and B (i.e., around region~4 marked in the panels) shows high velocity
dispersions ($\sim1500\kms$). The reason is that the multiple shocks
propagating in different directions significantly disturb the gas distributions
in these areas.

Panels~(d) in Figure~\ref{fig:xray_spec} show the X-ray spectra extracted from
the fields within the white squares indicated in the other panels of
Figure~\ref{fig:xray_spec}. The spectra are presented in the cluster rest
frame. For each model, we set four squares to cover the region in front of the
SE shock (region~1), the brightest X-ray emission core (region~2), the
wake-like structure (region~3), and the high velocity dispersion region
(region~4), respectively. The Doppler shifting and broadening of the spectral
lines are expected to contain the information on the line-of-sight motion of
the gas in the ICM. From the panels, we find the following main points.
\begin{itemize} \item In both of models A and B, the lines emitted from
regions~3 and 4 are significantly broadened, which includes the effects of the
thermal broadening and the Doppler broadening (also known as the kinetic
broadening). The relative contribution of these two mechanisms to the line
broadening can be approximated by
      \be \frac{W_{\rm therm}}{W_{\rm kin}}\simeq\frac{\sqrt{k_{\rm B}T_{\rm
gas}/5\keV}}{\sigma_{||}/10^2\kms}, \ee
      where $W_{\rm therm}$ and $W_{\rm kin}$ are the FWHMs of the thermal and
the kinetic broadenings, respectively (see eqs.~8, 11--12 in
\citealt{Kitayama2014}). The typical velocity dispersion and the gas
temperature in regions~3 and 4 are $\sim10^3\kms$ and $\sim30\keV$, respectively.
So the kinetic mechanism plays the dominative role in broadening the spectral
lines in ACT-CL J0102--4915. For example, the FWHM of the H-like Fe-K line emitted
from the twin-tailed structure of the cluster (in region~3, see Fig.~\ref{fig:xray_spec})
is around $\rm 50\,eV$. In addition, we do not see remarkable
He-like Fe-K lines emission from regions~3 and 4 in both of the models,
because of the generally high gas temperature in these areas.
  \item In fiducial model A, we can see a significant line shift ($\Delta
E\sim25{\rm\,eV}$) between the H-like Fe-K lines emitted from region~2 and
region~3, which approximately corresponds to $1000\kms$ in velocity. This
shift is mainly caused by the relative radial velocity between the primary and
the secondary clusters in model A. However, in model B, we do not find a
similar feature. The reason is that, the elongated gas distribution of the
secondary cluster in model B covers most of regions~2 and 3, which
holds a similar bulk velocity. The high-resolution spectroscopic measurement
in the future are expected to detect the line shifts among the different regions in
ACT-CL J0102--4915, and further provide a possible constraint on the merger
configuration of the cluster.  \end{itemize}

The Soft X-ray Spectrometer (SXS) on \emph{XARM} can provide an energy resolution of
$<7\eV$, which is expected to detect the modeled line features shown in
Figure~\ref{fig:xray_spec}(d) \citep[see][]{Takahashi2012}. Based on our
simulations, we simply estimate the required exposure time to identify the Fe-K lines near the
cluster X-ray center by using the properties of SXS instrument on
{\emph{ASTRO-H}} (i.e., angular resolution $\sim1.3'$, effective area $\sim200{\,\rm cm}^2$).
It approximately needs $\sim200{\rm\,ks}$ to collect $100$ photons for each Fe-K line (i.e.,
the peaks shown in the blue lines in panels (d) of Fig.~\ref{fig:xray_spec}).
The \emph{ATHENA} would further provide high spatial resolution measurements but also with
more than 10 times larger effective area \citep{Nandra2013}.
However, it is worthy of noting that we assume a universal metallicity $Z=0.3Z_{\odot}$
for both of the models in this study. But in reality, ACT-CL J0102--4915 reveals a higher
central metal abundance of $0.57\pm0.20$ \citep{Menanteau2012}. This spatially non-uniform
metal distribution may have impacts on the shape and the amplitude of the X-ray spectral lines.

\section{Conclusions} \label{sec:conclusions}

In this work, we predict the kSZ effect, relativistic correction of the tSZ effect,
and the X-ray spectra and line emission
of ACT-CL J0102--4915, based on fiducial models A and B proposed in ZYL15. The
kSZ effect and the broadening and the shifting of the spectral lines are
expected to provide insights into the gas motions of the ICM, which include the
bulk velocities of the merging subclusters and the turbulence produced by the
merger shocks. Our findings are concluded as follows.
\begin{itemize}
\item The amplitudes of the kSZ effect resulting from
fiducial models A and B are both at an order of $\Delta T/T\sim10^{-5}$
(the kSZ effect induced by the peculiar velocity of the entire cluster is
not included). The morphologies of the
kSZ effect resulting from these two models are different, and the difference
indicate the different line-of-sight velocity distributions in these two
models. Model A shows an almost axisymmetric morphology of the kSZ effect;
but model B shows a twisted one because of the off-axis merger
configuration. The maximal strength of the kSZ effect of model B
is roughly three times higher than that of model A. The kSZ measurements
thus provide a unique way to figure out the merger scenario of ACT-CL
J0102--4915. Though the kSZ effect is as weak as $\Delta T/T\sim10^{-5}$,
we find that combining the SZ distributions at
$148\ghz$ and $218\ghz$ could robustly extract the kSZ signals from the strong
tSZ emissions. The CMB instruments reaching $\rm\mu K$ sensitivity may detect
the kSZ signals of ACT-CL J0102--4915 in the future.
\item Both models A and B produce negative SZ temperature deviations around the
shock regions at $\sim230-250\ghz$, due to the relativistic correction of
the tSZ effect. These can be used to detect the possible high temperature
electrons heated by the shocks during the cluster merger, which is important
to give a constraint on the timescale of the thermal equilibrium between
the ions and the electrons.
\item Both models A and B reveal the shifting and broadening of the X-ray
spectral lines, which could be used to determine the velocity fields of the
merging cluster. The FWHM of the H-like Fe-K line emitted from the twin-tailed
structure of the cluster (in region~3, see Fig.~\ref{fig:xray_spec}) is around
$\rm 50\,eV$, mainly caused by the kinetic broadening mechanism.
The shifting of the X-ray spectral lines caused by the gas motion is different
in the two models, which can be up to $25\eV$ in some cluster regions.
In fiducial model A, we see a line shift between those emitted from the
brightest X-ray emission core (region~2, see Fig.~\ref{fig:xray_spec}) and from
the twin-tailed structure. However, we do not see the same feature in model B,
because the tidally elongated gas distribution of the secondary cluster in
model B covers most part of the sampling regions.
\end{itemize}

We conclude that the future observational instruments for the kSZ effect (e.g., ACT,
ALMA) or the X-ray spectral lines (e.g., \emph{XARM}, \emph{ATHENA})
may be able to detect the gas motions in ACT-CL J0102--4915, and give a constraint
on the cluster merger configurations.

We thank Daisuke Nagai, Erwin Lau, David Spergel, Rashid Sunyaev, and Eugene Churazov for suggestions
and discussions.  This work was supported in part by the National Natural Science
Foundation of China under nos.\ 11673001, 11273004, 11373031, 11390372, the
National Key R\&D Program of China (Grant Nos.  2016YFA0400703, 2016YFA0400704).


\begin{thebibliography}{0}
\expandafter\ifx\csname natexlab\endcsname\relax\def\natexlab#1{#1}\fi

\bibitem[Anders \& Grevesse(1989)]{Anders1989} Anders, E.,
\& Grevesse, N.\ 1989, \gca, 53, 197

\bibitem[Basu et al.(2016)]{Basu2016} Basu, K., Sommer, M., Erler, J., et al.\ 2016, \apjl, 829, L23

\bibitem[Benson et al.(2014)]{Benson2014} Benson, B.~A., Ade, P.~A.~R.,
Ahmed, Z., et al.\ 2014, \procspie, 9153, 91531P

\bibitem[Biffi et al.(2013)]{Biffi2013} Biffi, V., Dolag, K.,
B\"{o}hringer, H.\ 2013, \mnras, 428, 1395

\bibitem[Birkinshaw(1999)]{Birkinshaw1999} Birkinshaw, M.\ 1999,
\physrep, 310, 97

\bibitem[Fryxell et~al.(2000) ]{Fryxell2000} Fryxell, B., Olson, K.,
Ricker, P., et~al.\ 2000, \apjs, 131, 273

\bibitem[Hand et al.(2012)]{Hand2012} Hand, N., Addison, G.~E.,
Aubourg, E., et al.\ 2012, Physical Review Letters, 109, 041101

\bibitem[Itoh et~al.(1998)]{Itoh1998} {Itoh}, N., {Kohyama}, Y., \&
{Nozawa}, S.\ 1998, \apj, 502, 7

\bibitem[Jee et al.(2014)]{Jee2014} Jee, M.~J., Hughes, J.~P.,
Menanteau, F., et al.\ 2014, \apj, 785, 20

\bibitem[Kitayama et al.(2014)]{Kitayama2014} Kitayama, T., Bautz,
M., Markevitch, M., et al.\ 2014, arXiv:1412.1176

\bibitem[Lindner et al.(2014)]{Lindner2014} Lindner, R.~R., Baker,
A.~J., Hughes, J.~P., et al.\ 2014, \apj, 786, 49

\bibitem[Lindner et al.(2015)]{Lindner2015} Lindner, R.~R., Aguirre,
P., Baker, A.~J., et al.\ 2015, \apj, 803, 79

\bibitem[Marriage et al.(2011)]{Marriage2011} Marriage, T.~A.,
Acquaviva, V., Ade, P.~A.~R., et al.\ 2011, \apj, 737, 61

\bibitem[Mazzotta et al.(2004)]{Mazzotta2004} Mazzotta, P.,
Rasia, E., Moscardini, L., \& Tormen, G.\ 2004, \mnras, 354, 10

\bibitem[Menanteau et al.(2012)]{Menanteau2012} Menanteau, F.,
Hughes, J.~P., Sif{\'o}n, C., et al.\ 2012, \apj, 748, 7

\bibitem[Mroczkowski et al.(2012)]{Mroczkowski2012} Mroczkowski, T.,
Dicker, S., Sayers, J., et al.\ 2012, \apj, 761, 47

\bibitem[Nandra et al.(2013)]{Nandra2013} Nandra, K., Barret, D., Barcons, X., et al.\ 2013, arXiv:1306.2307

\bibitem[Nozawa et al.(1998)]{Nozawa1998} Nozawa, S., Itoh, N.,
\& Kohyama, Y.\ 1998, \apj, 508, 17

\bibitem[Nozawa et al.(2000)]{Nozawa2000} Nozawa, S., Itoh, N., Kawana, Y.,
\& Kohyama, Y.\ 2000, \apj, 536, 31

\bibitem[Planck Collaboration et al.(2016)]{Planck2016_I37} Planck Collaboration, Ade,
P.~A.~R., Aghanim, N., et al.\ 2016, \aap, 586, A140

\bibitem[Ricker(2008)]{Ricker2008} Ricker, P.~M. 2008, \apjs, 176, 293

\bibitem[Rubi{\~n}o-Mart{\'{\i}}n et al.(2004)]{Martin2004} Rubi{\~n}o-Mart{\'{\i}}n, J.~A.,
Hern{\'a}ndez-Monteagudo, C., \& En{\ss}lin, T.~A.\ 2004, \aap, 419, 439

\bibitem[Russell et al.(2012)]{Russell2012} Russell, H.~R., McNamara,
B.~R., Sanders, J.~S., et al.\ 2012, \mnras, 423, 236

\bibitem[Sayers et al.(2013)]{Sayers2013} Sayers, J., Mroczkowski, T.,
Zemcov, M., et al.\ 2013, \apj, 778, 52

\bibitem[Sayers et al.(2016)]{Sayers2016} Sayers, J., Zemcov, M.,
Glenn, J., et al.\ 2016, \apj, 820, 101

\bibitem[Schaan et al.(2015)]{Schaan2015} Schaan, E., Ferraro, S.,
Vargas-Maga{\~n}a, M., et al.\ 2015, arXiv:1510.06442

\bibitem[Soergel et al.(2016)]{Soergel2016} Soergel, B., Flender, S.,
Story, K.~T., et al.\ 2016, arXiv:1603.03904

\bibitem[Springel et~al.(2001)]{Springel2001} {Springel}, V., {Yoshida}, N.,
\& {White}, S.~D.~M.\ 2001, \na, 6, 79

\bibitem[Sunyaev et al.(2003)]{Sunyaev2003} Sunyaev, R.~A., Norman, M.~L.,
\& Bryan, G.~L.\ 2003, Astronomy Letters, 29, 783

\bibitem[Takahashi et al.(2012)]{Takahashi2012} Takahashi, T., Mitsuda, K., Kelley, R., et al.\ 2012, \procspie, 8443, 84431Z

\bibitem[Yamada et al.(2012)]{Yamada2012} Yamada, K., Kitayama, T., Takakuwa, S., et al.\ 2012, \pasj, 64, 102

\bibitem[Zhang et al.(2015)]{Zhang2015} Zhang, C., Yu, Q.,
\& Lu, Y.\ 2015, \apj, 813, 129



\end{thebibliography}
\end{document}